\documentclass[11pt]{article}
\usepackage[latin9]{inputenc}
\usepackage{amsfonts}
\usepackage{amsmath}
\usepackage{amssymb}
\usepackage{indentfirst}
\usepackage{graphicx}
\usepackage[colorlinks]{hyperref}
\usepackage{cite}

\numberwithin{equation}{section}
\begin{document}

\baselineskip=18pt 
\baselineskip 0.6cm
\begin{titlepage}
\vskip 4cm

\begin{center}
\textbf{\LARGE{Three-dimensional Maxwellian extended Bargmann supergravity}}
\par\end{center}{\LARGE \par}

\begin{center}
	\vspace{1cm}
	\textbf{Patrick Concha}$^{\ast}$,
    \textbf{Lucrezia Ravera}$^{\ddag}$,
	\textbf{Evelyn Rodríguez}$^{\dag}$,
	\small
	\\[6mm]
	$^{\ast}$\textit{Departamento de Matemática y Física Aplicadas, }\\
	\textit{ Universidad Católica de la Santísima Concepción, }\\
\textit{ Alonso de Ribera 2850, Concepción, Chile.}
	\\[3mm]
    $^{\ddag}$\textit{INFN, Sezione di Milano, }\\
	\textit{ Via Celoria 16, I-20133 Milano-Italy.}
	\\[3mm]
	$^{\dag}$\textit{Departamento de Ciencias, Facultad de Artes Liberales,} \\
	\textit{Universidad Adolfo Ibáñez, Viña del Mar-Chile.} \\[5mm]
	\footnotesize
	\texttt{patrick.concha@ucsc.cl},
    \texttt{lucrezia.ravera@mi.infn.it},
	\texttt{evelyn.rodriguez@edu.uai.cl},
	\par\end{center}
\vskip 20pt
\centerline{{\bf Abstract}}
\medskip
\noindent We present a novel three-dimensional non-relativistic Chern-Simons supergravity theory invariant under a Maxwellian extended Bargmann superalgebra. We {first study} the non-relativistic limits of the minimal and {the} $\mathcal{N}=2$ Maxwell superalgebra{s}. We show that a well-defined Maxwellian extended Bargmann supergravity requires to construct by hand a supersymmetric extension of the Maxwellian extended Bargmann algebra by introducing additional fermionic and bosonic generators. The new non-relativistic supergravity action presented here contains the extended Bargmann supergravity as a sub-case.

\end{titlepage}\newpage {}

\section{Introduction}

Three-dimensional non-relativistic (NR) and ultra-relativistic (UR) versions
of supergravity theory {have} only been explored recently in \cite{ABRS,
BRZ, BR, OOTZ, OOZ, Rav, FR}. Although several generalizations and
applications of supergravity have been developed by diverse authors these
last four decades, its NR construction remains challenging and has only been
approached in three spacetime dimensions. In particular, the formulation of
a well-defined NR supergravity action has required the introduction of
additional fermionic generators. As in NR bosonic cases, the addition of new
generators allows to construct a non-degenerate invariant bilinear form
which assures the proper construction of a Chern-Simons (CS) action.

The NR theories {have} received a renewed interest since they play an
important role to approach condensed matter systems \cite{Son, BM, KLM, BG,
BGMM, CHOR, CHOR2, Taylor} and NR effective field theories \cite{Son2, HS,
GPR, GJA}. It seems then natural to extend NR gravity theories \cite{DK,
DBKP, DGH, Duval, DLP, DLP2, Horava, DH, PS, ABPR, ABGR, BMM, BM1, HLO, BCRR,
BM2, CS} to the presence of supersymmetry. In particular, NR supergravity
models can be seen as a starting point to approach supersymmetric field
theories on curved backgrounds by means of localization \cite{FS, Pestun}.

On the other hand, the Maxwell algebra has received a growing interest these
last decades. Such symmetry has been {first introduced} to describe
Minkowski space in the presence of a constant electromagnetic field
background \cite{Schrader, BCR, GK}. In the gravity context, the Maxwell
algebra and its generalizations have been useful to recover standard General
Relativity from CS and Born-Infeld gravity theories \cite{EHTZ, GRCS, CPRS1,
CPRS2, CPRS3}. More recently, a Maxwell CS formulation in three spacetime
dimensions has been explored in \cite{SSV}. Its solution \cite{HR, CMMRSV},
generalization to higher spin \cite{CCFRS}, and asymptotic symmetry \cite%
{CMMRSV, CCRS} have been subsequently studied by diverse authors. Further
application of the Maxwell algebra can be found in \cite{AKL, DKGS, AKL2,
CK, BS, KSC, SalgadoReb}. At the supersymmetric level, the minimal Maxwell
superalgebra appears to describe a constant Abelian supersymmetric gauge
field background in a four-dimensional superspace \cite{BGKL}.
Generalizations of the Maxwell superalgebras have then been explored with
diverse applications \cite{BGKL2, Lukierski, AILW, AI, CR1, CR2, PR, Ravera,
CRR, KC}.\ More recently, a three-dimensional CS supergravity theory
invariant under the Maxwell superalgebra and its $\mathcal{N}$-extended
versions have been explored in \cite{CFRS, CFR, CPR, Concha}.

The NR version of the Maxwell CS gravity theory {has} only been presented
recently \cite{AFGHZ} (see also \cite{GKPSR}, where the related algebra has
been recovered through Lie algebra expansion). Interestingly, the
relativistic theory required the presence of three U(1) gauge fields in
order to establish a well-defined NR limit and to avoid degeneracy. In the
presence of supersymmetry, the NR version of the Maxwell CS supergravity was
unknown till now. In this work, we explore the NR limit of the Maxwell
superalgebra for $\mathcal{N}=1$ and $\mathcal{N}=2$. In particular, we show
that a well-defined NR Maxwellian CS supergravity action requires to
introduce by hand additional fermionic and bosonic generators. Our model is
not only a novel NR supergravity theory without cosmological constant but
contains the extended Bargmann supergravity as a sub-case.

The paper is organized as follows: In Section 2, we briefly review the
Maxwellian extended Bargmann gravity introduced in \cite{AFGHZ}. Sections 3
and 4 contain our main results. In Section 3, we introduce the NR limits of
the minimal and the $\mathcal{N}=2$ Maxwell superalgebras. In Section 4, we
present the Maxwellian extendad Bargmann superalgebra and the NR CS
supergravity action. Section 5 is devoted to discussion and possible future
developments. {Some large formulas are collected in the Appendix.}

\section{Maxwellian extended Bargmann gravity}

In this section, we briefly review the Maxwellian {e}xtended Bargmann
algebra introduced in \cite{AFGHZ} and the associated {CS} gravity theory
developed in the same paper in three (2+1) dimensions. In \cite{AFGHZ} the
authors proved that an alternative way to circumvent the degeneracy of the
bilinear form in the [Maxwell] $\oplus$ $u(1)$ $\oplus$ $u(1)$ system
analyzed in the same paper, is to add one more $u(1)$ gauge field.

The {non-vanishing} commutation relations of the Maxwell algebra are given
by
\begin{eqnarray}
\left[ J_{A},J_{B}\right] &=&\epsilon _{ABC}J^{C}\,,  \notag \\
\left[ J_{A},P_{B}\right] &=&\epsilon _{ABC}P^{C}\,,  \notag \\
\left[ J_{A},Z_{B}\right] &=&\epsilon _{ABC}Z^{C}\,,  \label{m1} \\
\left[ P_{A},P_{B}\right] &=&\epsilon _{ABC}Z^{C}\,,  \notag
\end{eqnarray}
where $J_A$ are the spacetime rotations, $P_A$ the spacetime translations,
and $Z_A$ are new generators characterized and introduced in \cite{Schrader,
BCR} ($A=0,1,2$ and $\eta^{AB}=\text{diag}(-,+,+)$). A gauge-invariant {CS}
gravity action in three (meaning 2+1, here as well as in the sequel)
dimensions based on the above written Maxwell algebra {has been constructed
in \cite{SSV, HR, AFGHZ, CMMRSV}. The CS action is constructed} using the
connection one-form $A = A^A T_A$ taking values in the Maxwell algebra
generated by $\lbrace J_A, P_A, Z_A \rbrace$, that is
\begin{equation}
A = E^B P_B +W^B J_B + K^B Z_B \, ,
\end{equation}
where $E^B$, $W^B$, and $K^B$ are one-form fields.

{The CS form} constructed with the invariant bilinear form defines an action
for the relativistic gauge theory for the symmetry under consideration as
\begin{equation}  \label{genCS}
I_{\text{CS}} = \int \langle A \wedge dA + \frac{2}{3} A \wedge A \wedge A
\rangle = \int \langle A \wedge dA + \frac{1}{3} A \wedge \left[ A , A %
\right] \rangle \, .
\end{equation}

In the specific case we are now reviewing, when the Maxwell algebra is
supplemented with the three additional $U(1)$ generators ($Y_1$, $Y_2$, and $%
Y_3$), the connection {one-form} involved in the construction reads
\begin{equation}
A = E^B P_B +W^B J_B + K^B Z_B + M Y_1 + S Y_2 + T Y_3 \, ,
\end{equation}
where $M$, $S$, and $T$ are the additional bosonic {gauge} fields. Also the
bilinear form acquires further non-zero entries due to the presence of the
new generators (see \cite{AFGHZ} for details). In particular, a
non-degenerate bilinear form can be obtained from the aforesaid relativistic
bilinear form, allowing for a well-defined and finite NR CS action.

Specifically, in \cite{AFGHZ}, the contraction leading to the NR {generators}
is defined through the identifications
\begin{eqnarray}
P_{0} &=&\frac{\tilde{H}}{2\xi }+\xi \tilde{M}\,,\text{ \ \ \ }P_{a}=\tilde{P%
}_{a}\,,\text{ \ \ \ \ \ \ \ }Y_1=\frac{\tilde{H}}{2\xi }-\xi \tilde{M}\,,
\notag \\
J_{0} &=&\frac{\tilde{J}}{2}+ \xi^2 \tilde{S} \,,\text{ \ \ \ \ }J_{a}=\xi
\tilde{G}_a \,,\text{ \ \ \ \ \ \ }Y_2=\frac{\tilde{J}}{2}-\xi^2 \tilde{S}\,,
\label{contr1} \\
Z_{0} &=&\frac{\tilde{Z}}{2 \xi^2}+ \tilde{T} \,,\text{ \ \ \ \ }Z_{a}=\frac{%
\tilde{Z}_{a}}{\xi }\,,\text{ \ \ \ \ \ \ }Y_3=\frac{\tilde{Z}}{2\xi ^{2}}-%
\tilde{T}\,,  \notag
\end{eqnarray}
{and by subsequently taking $\xi \rightarrow \infty$. Let us note that} the
index $A=0,1,2$ has previously been decomposed as $A \rightarrow \lbrace 0,
a \rbrace$, with $a=1,2$. {Furthermore,} $Y_1$, $Y_2$, and $Y_3$ are the
three $U(1)$ generators introduced at the relativistic level.

In terms of the NR generators and fields, the {gauge connection one-form} of
\cite{AFGHZ}, $\tilde{A} = A^A \tilde{T}_A$, is given by
\begin{equation}
\tilde{A} = \tau \tilde{H} + e^a \tilde{P}_a + \omega \tilde{J} + \omega^a
\tilde{G}_a + k \tilde{Z} + k^a \tilde{Z}_a + m \tilde{M} + s \tilde{S} + t
\tilde{T} \, .
\end{equation}
The {NR version of the} Maxwell algebra presented in \cite{AFGHZ} was called
by the authors Maxwellian Exotic Bargmann (MEB) algebra, and its non-trivial
commutations relations read
\begin{eqnarray}
\left[ \tilde{G}_{a},\tilde{P}_{b}\right] &=&-\epsilon _{ab}\tilde{M}%
\,,\qquad \left[ \tilde{G}_{a},\tilde{Z}_{b}\right] =-\epsilon _{ab}\tilde{T}%
\,,\text{ \ }  \notag \\
\left[ \tilde{H},\tilde{G}_{a}\right] &=&\epsilon _{ab}\tilde{P}%
_{b}\,,\qquad \quad \ \left[ \tilde{J},\tilde{Z}_{a}\right] =\epsilon _{ab}%
\tilde{Z}_{b}\,,  \notag \\
\left[ \tilde{J},\tilde{P}_{a}\right] &=&\epsilon _{ab}\tilde{P}%
_{b}\,,\qquad \quad \left[ \tilde{H},\tilde{P}_{a}\right] =\epsilon _{ab}%
\tilde{Z}_{b}\,,  \label{MEB1} \\
\left[ \tilde{J},\tilde{G}_{a}\right] &=&\epsilon _{ab}\tilde{G}%
_{b}\,,\qquad \ \ \left[ \tilde{P}_{a},\tilde{P}_{b}\right] =-\epsilon _{ab}%
\tilde{T}\,,  \notag \\
\left[ \tilde{G}_{a},\tilde{G}_{b}\right] &=&-\epsilon _{ab}\tilde{S}%
\,,\qquad \quad \ \left[ \tilde{Z},\tilde{G}_{a}\right] =\epsilon _{ab}%
\tilde{Z}_{b}\,.  \notag
\end{eqnarray}
Such NR algebra admits the following non-vanishing components of the
invariant tensor 
\begin{eqnarray}
\left\langle \tilde{G}_a \tilde{G}_b \right\rangle &=& \tilde{\alpha}_0
\delta_{ab} \,,  \notag \\
\left\langle \tilde{G}_a \tilde{P}_b \right\rangle &=& \tilde{\alpha}_1
\delta_{ab} \,,  \notag \\
\left\langle \tilde{G}_a \tilde{Z}_b \right\rangle &=& \tilde{\alpha}_2
\delta_{ab} \, = \, \left\langle \tilde{P}_a \tilde{P}_b \right\rangle \,,  \label{invt1} \\
\left\langle \tilde{J} \tilde{S} \right\rangle &=& -\tilde{\alpha}_0 \,,
\notag \\
\left\langle \tilde{J} \tilde{M} \right\rangle &=& -\tilde{\alpha}_1 \, = \,
\left\langle \tilde{H} \tilde{S} \right\rangle \,,  \notag \\
\left\langle \tilde{J} \tilde{T} \right\rangle &=& -\tilde{\alpha}_2 \, = \,
\left\langle \tilde{H} \tilde{M} \right\rangle \,.  \notag
\end{eqnarray}
This bilinear form is non-degenerate if $\tilde{\alpha}_2 \neq 0$. The MEB
curvature two-forms are given by
\begin{eqnarray}
R\left( \omega \right) &=&d\omega \,,  \notag \\
R^{a}\left( \omega ^{b}\right) &=&d\omega ^{a}+\epsilon ^{ac}\omega
\omega_{c}\,,  \notag \\
R\left( \tau \right) &=& d\tau \,,  \notag \\
R^{a}\left( e^{b}\right) &=& d e^a + \epsilon ^{ac}\omega e_{c} + \epsilon
^{ac}\tau \omega_{c} \,,  \notag \\
R\left( k\right) &=& dk \,,  \label{curvMEB} \\
R^{a}\left( k^{b}\right) &=& d k^a + \epsilon ^{ac}\omega k_{c} + \epsilon
^{ac}\tau e_{c} + \epsilon ^{ac} k \omega_{c} \,,  \notag \\
R\left( m\right) &=& dm + \epsilon ^{ac} e_a \omega_{c} \,,  \notag \\
R\left( s\right) &=&ds+\frac{1}{2}\epsilon ^{ac}\omega _{a}\omega _{c}\,,
\notag \\
R\left( t\right) &=& dt + \epsilon ^{ac}\omega_{a} k_{c} + \frac{1}{2}
\epsilon ^{ac} e_{a} e_{c} \,.  \notag
\end{eqnarray}

The NR three-dimensional CS action obtained in \cite{AFGHZ} reads, up to
boundary terms, as follows:
\begin{eqnarray}
I_{\text{MEB}} &=& \int \Bigg \lbrace \tilde{\alpha}_{0}\bigg[ \omega
_{a}R^{a}(\omega^{b})-2sR\left( \omega \right) \bigg] +\tilde{\alpha}_{1}%
\bigg[ 2e_{a}R^{a}(\omega ^{b})-2mR(\omega )-2\tau R(s) \bigg]  \notag \\
&& +\tilde{\alpha}_{2} \bigg[ e_{a}R^{a}\left( e^{b}\right)
+k_{a}R^{a}\left( \omega ^{b}\right) +\omega _{a}R^{a}\left( k^{b}\right)
-2sR\left( k\right) -2mR\left( \tau \right)  \notag \\
&& - 2tR\left( \omega \right) \bigg] \Bigg \rbrace \,.  \label{CS1}
\end{eqnarray}
As was noticed in \cite{AFGHZ}, the NR CS action has three independent
sectors proportional to three arbitrary constants, $\tilde{\alpha}_0$, $%
\tilde{\alpha}_1$, and $\tilde{\alpha}_2$. The first term corresponds to the
so-called exotic NR gravity. The second term is the CS action for the
extended Bargmann algebra \cite{LL, Grigore, Bose, DHO2, JN, HP}, while the
last term reproduces the CS action for a new NR Maxwell algebra. Let us note
that, since the bilinear form does not result to acquire degeneracy in the
contraction process, {the equations of motion from the NR action (\ref{CS1})
are given by the vanishing of all the curvatures (\ref{curvMEB}).}

\section{On the supersymmetric extension of the Maxwellian {e}xtended
Bargmann algebra}

In this section{,} we explore the supersymmetric extension of the NR Maxwell
algebra by applying a NR limit to the $\mathcal{N}=1$ and $\mathcal{N}=2$
Maxwell superalgebra. Interestingly{,} we show that, in order to have a
well-defined NR superalgebra, we have to consider the NR limit of a
centrally extended $\mathcal{N}=2$ Maxwell superalgebra endowed with a $%
\mathfrak{so}(2)$ generator. Indeed{,} a true supersymmetric {extension} of
the MEB algebra in which the anti-commutator of two fermionic charges gives
a time and a space translation requires, as in the Bargmann case, at least $%
\mathcal{N}=2$ supersymmetry. However, as we shall see in the next section,
it is necessary to introduce by hand additional fermionic and bosonic
generators in order to obtain a MEB\ superalgebra which allows the proper
construction of a NR CS supergravity.

In three spacetime dimensions, the minimal Maxwell superalgebra is spanned
by the set of generators $\left\{ J_{A},P_{A},Z_{A},Q_{\alpha },\Sigma
_{\alpha }\right\} $ \cite{CPR}{, where, in particular, $Q_\alpha$ are the supersymmetry generators. Besides, such} supersymmetric extension of the
Maxwell algebra is characterized by the presence of an additional Majorana
fermionic generator $\Sigma_\alpha$ whose presence assures the Jacobi
identity $\left( P_a,Q_\alpha,Q_\beta\right)$. The introduction of a second
spinorial charge is not new and have previously been considered in
superstring theory \cite{Green} and $D=11$ supergravity \cite{AF, AAR1, AAR2}. The (anti-)commutation relations of the minimal Maxwell superalgebra are
given by%
\begin{eqnarray}
\left[ J_{A},J_{B}\right] &=&\epsilon _{ABC}J^{C}\,,  \notag \\
\left[ J_{A},P_{B}\right] &=&\epsilon _{ABC}P^{C}\,,  \notag \\
\left[ J_{A},Z_{B}\right] &=&\epsilon _{ABC}Z^{C}\,,  \notag \\
\left[ P_{A},P_{B}\right] &=&\epsilon _{ABC}Z^{C}\,,  \notag \\
\left[ J_{A},Q_{\alpha }\right] &=&-\frac{1}{2}\,\left( \gamma _{A}\right)
_{\alpha }^{\text{ }\beta }Q_{\beta }\, , \text{ \ \ }  \label{sm1} \\
\left[ J_{A},\Sigma _{\alpha }\right] &=&-\frac{1}{2}\,\left( \gamma
_{A}\right) _{\alpha }^{\text{ }\beta }\Sigma _{\beta }\,,  \notag \\
\left[ P_{A},Q_{\alpha }\right] &=&-\frac{1}{2}\,\left( \gamma _{A}\right)
_{\alpha }^{\text{ }\beta }\Sigma _{\beta }\,,\text{ }  \notag \\
\left\{ Q_{\alpha },Q_{\beta }\right\} &=&-\left( \gamma ^{A}C\right)
_{\alpha \beta }P_{A}\,,  \notag \\
\left\{ Q_{\alpha },\Sigma _{\beta }\right\} &=&-\,\left( \gamma
^{A}C\right) _{\alpha \beta }Z_{A}\,,  \notag
\end{eqnarray}%
where $\alpha {, \beta} =1,2$ are spinorial indices, $C$ is the charge conjugation
matrix, and $\gamma ^{A}$ are the Dirac matrices in three spacetime
dimensions.

As was discussed in \cite{AFGHZ}, it is necessary to include\ three
additional $U\left( 1\right) $ generators given by $Y_{1}$, $Y_{2}$, and $%
Y_{3}$ in order to get the bosonic MEB algebra as a NR limit. At the
supersymmetric level, a NR contraction can be applied by considering the
rescaling of the bosonic generators as in (\ref{contr1}) and the following
rescaling, with a dimensionless parameter $\xi $, of the {Majorana} fermionic
generators {$Q_\alpha$ and $\Sigma_\alpha$}:
\begin{equation}
Q_{{\alpha}}=\sqrt{\xi }\tilde{Q}_{{\alpha}}^{-}\,,\qquad \Sigma_{{\alpha}} =\frac{1}{\sqrt{\xi}} \tilde{\Sigma}_{{\alpha}}^{-} \,.
\end{equation}
A particular supersymmetric extension of the MEB\ algebra is obtained from
the NR contraction $\xi \rightarrow \infty $ of (\ref{sm1}):

\begin{eqnarray}
\left[ \tilde{G}_{a},\tilde{P}_{b}\right] &=&-\epsilon _{ab}\tilde{M}%
\,,\qquad \ \ \ \ \ \ \ \ \left[ \tilde{G}_{a},\tilde{Z}_{b}\right]
=-\epsilon _{ab}\tilde{T}\,,\text{ \ }  \notag \\
\left[ \tilde{H},\tilde{G}_{a}\right] &=&\epsilon _{ab}\tilde{P}%
_{b}\,,\qquad \quad \ \ \ \ \ \ \ \ \ \left[ \tilde{J},\tilde{Z}_{a}\right]
=\epsilon _{ab}\tilde{Z}_{b}\,,  \notag \\
\left[ \tilde{J},\tilde{P}_{a}\right] &=&\epsilon _{ab}\tilde{P}%
_{b}\,,\qquad \quad \ \ \ \ \ \ \ \ \left[ \tilde{H},\tilde{P}_{a}\right]
=\epsilon _{ab}\tilde{Z}_{b}\,,  \notag \\
\left[ \tilde{J},\tilde{G}_{a}\right] &=&\epsilon _{ab}\tilde{G}%
_{b}\,,\qquad \ \ \ \ \ \ \ \ \ \ \left[ \tilde{P}_{a},\tilde{P}_{b}\right]
=-\epsilon _{ab}\tilde{T}\,,  \notag \\
\left[ \tilde{G}_{a},\tilde{G}_{b}\right] &=&-\epsilon _{ab}\tilde{S}%
\,,\qquad \quad \ \ \ \ \ \ \ \left[ \tilde{Z},\tilde{G}_{a}\right]
=\epsilon _{ab}\tilde{Z}_{b}\,,  \label{N1} \\
\left[ \tilde{J},\tilde{Q}_{\alpha }^{-}\right] &=&-\frac{1}{2}\left( \gamma
_{0}\right) _{\alpha }^{\text{ }\beta }\tilde{Q}_{\beta }^{-}\,,\text{ \ \ \
\ \ }\left[ \tilde{J},\tilde{\Sigma}_{\alpha }^{-}\right] =-\frac{1}{2}%
\left( \gamma _{0}\right) _{\alpha }^{\text{ }\beta }\tilde{\Sigma}_{\beta
}^{-}\,,  \notag \\
\left[ \tilde{H},\tilde{Q}_{\alpha }^{-}\right] &=&-\frac{1}{2}\left( \gamma
_{0}\right) _{\alpha }^{\text{ }\beta }\tilde{\Sigma}_{\beta }^{-}\,,  \notag
\\
\left\{ \tilde{Q}_{\alpha }^{-},\tilde{Q}_{\beta }^{-}\right\} &=&-\left(
\gamma ^{0}C\right) _{\alpha \beta }\tilde{M}\,,\quad \left\{ \tilde{Q}%
_{\alpha }^{-},\tilde{\Sigma}_{\beta }^{-}\right\} =-\left( \gamma
^{0}C\right) _{\alpha \beta }\tilde{T}\,.  \notag
\end{eqnarray}%
Although the (anti-)commutation relations (\ref{N1}) are well-defined and
satisfy the Jacobi identities, we cannot say that the $\mathcal{N}=1$ MEB
superalgebra obtained here is a true supersymmetry algebra. Indeed, the
anti-commutator of two supercharges leads to a central charge transformation
instead of a time and space translation. This is analogous to the $\mathcal{N%
}=1$ Bargmann superalgebra case \cite{ABRS}.

One way to circumvent such difficulty is to apply the NR contraction to a $%
\mathcal{N}=2$ relativistic Maxwell superalgebra. The $\mathcal{N}=2$
supersymmetric extension of the Maxwell algebra has been explored by diverse
authors \cite{AILW, CR1, CFR}. Here, we shall focus on the $\mathcal{N}=2$
centrally extended Maxwell superalgebra endowed with a $\mathfrak{so}\left(
2\right) $ internal symmetry generator introduced in \cite{Concha}. Such $%
\mathcal{N}=2$ Maxwell superalgebra is spanned by the set of generators $%
\left\{ J_{A},P_{A},Z_{A},\mathcal{B},\mathcal{Z},Q_{\alpha }^{i},\Sigma
_{\alpha }^{i}\right\} ${,} which satisfy the following non-vanishing
(anti-)commutation relations:
{
\begin{eqnarray}
\left[ J_{A},J_{B}\right] &=&\epsilon _{ABC}J^{C}\,, \qquad \quad \ \
\left[J_{A},P_{B}\right] =\epsilon _{ABC}P^{C}\,,  \notag \\
\left[ J_{A},Z_{B}\right] &=&\epsilon _{ABC}Z^{C}\,,  \qquad \quad \ \
\left[P_{A},P_{B}\right] =\epsilon _{ABC}Z^{C}\,,  \notag \\
\left[ J_{A},Q_{\alpha }^{i}\right] &=&-\frac{1}{2}\,\left( \gamma
_{A}\right) _{\alpha }^{\text{ }\beta }Q_{\beta }^{i}\,,\quad \ \left[
J_{A},\Sigma _{\alpha }^{i}\right] =-\frac{1}{2}\,\left( \gamma _{A}\right)
_{\alpha }^{\text{ }\beta }\Sigma _{\beta }^{i}\,,  \label{N2Mnocs} \\
\left[ P_{A},Q_{\alpha }^{i}\right] &=&-\frac{1}{2}\,\left( \gamma
_{A}\right) _{\alpha }^{\text{ }\beta }\Sigma _{\beta }^{i}\,,  \quad \ \ \left[ Q_{\alpha }^{i},\mathcal{B}\right] = \frac{1}{2}\epsilon
^{ij}\Sigma _{\alpha }^{j}\,,  \notag \\
\left\{ Q_{\alpha }^{i},Q_{\beta }^{j}\right\} &=&-\delta ^{ij}\left( \gamma
^{A}C\right) _{\alpha \beta }P_{A}\,-C_{\alpha \beta }\epsilon ^{ij}\mathcal{%
B}\,,  \notag \\
\left\{ Q_{\alpha }^{i},\Sigma _{\beta }^{j}\right\} &=&-\delta
^{ij}\,\left( \gamma ^{A}C\right) _{\alpha \beta }Z_{A}\,-C_{\alpha \beta
}\epsilon ^{ij}\mathcal{Z}\, ,  \notag
\end{eqnarray}}%
where $i=1,2$ is the number of supercharges. Let us note that the presence
of a $\mathfrak{so}\left( 2\right) $ internal symmetry generator is crucial
in order to admit a non-degenerate invariant inner product \cite{Concha}.
Then, following \cite{LPSZ}, let us consider the following definitions of
the fermionic generators%
\begin{eqnarray}
Q_{\alpha }^{\pm } &=&\frac{1}{\sqrt{2}}\left( Q_{\alpha }^{1}\pm \epsilon
_{\alpha \beta }Q_{\beta }^{2}\right) \,,  \notag \\
\Sigma _{\alpha }^{\pm } &=&\frac{1}{\sqrt{2}}\left( \Sigma _{\alpha
}^{1}\pm \epsilon _{\alpha \beta }\Sigma _{\beta }^{2}\right) \,.
\end{eqnarray}%
A dimensionless parameter $\xi $ can be introduced by considering the
rescaling of the generators and central extension,%
\begin{eqnarray}
J_{0} &=&\tilde{J}\,,\text{\ \ \ \quad \quad \quad\ \ \ }J_{a}=\xi \tilde{G}%
_{a}\,,\text{ \ \ \ \ \ \ },  \notag \\
P_{0} &=&\frac{\tilde{H}}{2\xi }+\xi \tilde{M}\,,\text{ \ \ \ }P_{a}=\tilde{P%
}_{a}\,,\text{ \ \ \ \ \ \ \ }\mathcal{B}=\frac{\tilde{H}}{2\xi }-\xi \tilde{%
M}\,,  \notag \\
Z_{0} &=&\frac{\tilde{Z}}{2\xi ^{2}}+\tilde{T}\,,\text{ \ \ \ \ }Z_{a}=\frac{%
\tilde{Z}_{a}}{\xi }\,,\text{ \ \ \ \ \ \ }\mathcal{Z}=\frac{\tilde{Z}}{2\xi
^{2}}-\tilde{T}\,, \\
Q_{\alpha }^{-} &=&\sqrt{\xi }\tilde{Q}_{\alpha }^{-}\,,\quad \quad
Q_{\alpha }^{+}=\frac{1}{\sqrt{\xi }}\tilde{Q}_{\alpha }^{+}\,,  \notag \\
\Sigma _{\alpha }^{-} &=&\frac{1}{\sqrt{\xi }}\tilde{\Sigma}_{\alpha
}^{-}\,,\qquad \Sigma _{\alpha }^{+}=\frac{1}{\xi ^{3/2}}\tilde{\Sigma}%
_{\alpha }^{+}\,.  \notag
\end{eqnarray}%
Then, after taking the limit $\xi \rightarrow \infty $, a particular $%
\mathcal{N}=2$ Maxwellian Bargmann superalgebra is obtained; its
(anti-)commutation relations are given by the purely bosonic commutators%
\begin{eqnarray}
\left[ \tilde{G}_{a},\tilde{P}_{b}\right] &=&-\epsilon _{ab}\tilde{M}\,,%
\text{ \ \ \ \ \ \ }\left[ \tilde{G}_{a},\tilde{Z}_{b}\right] =-\epsilon
_{ab}\tilde{T}\,,\text{ \ }  \notag \\
\left[ \tilde{H},\tilde{G}_{a}\right] &=&\epsilon _{ab}\tilde{P}_{b}\,,\text{
\ \ \ \ \ \ \ \ \ \ }\left[ \tilde{J},\tilde{Z}_{a}\right] =\epsilon _{ab}%
\tilde{Z}_{b}\,,  \notag \\
\left[ \tilde{J},\tilde{P}_{a}\right] &=&\epsilon _{ab}\tilde{P}_{b}\,,\text{
\ \ \ \ \ \ \ \ \ }\left[ \tilde{H},\tilde{P}_{a}\right] =\epsilon _{ab}%
\tilde{Z}_{b}\,,  \label{MB} \\
\left[ \tilde{J},\tilde{G}_{a}\right] &=&\epsilon _{ab}\tilde{G}_{b}\,,\text{
\ \ \ \ \ \ \ \ }\left[ \tilde{P}_{a},\tilde{P}_{b}\right] =-\epsilon _{ab}%
\tilde{T}\,,  \notag \\
\left[ \tilde{Z},\tilde{G}_{a}\right] &=&\epsilon _{ab}\tilde{Z}_{b}\,,
\notag
\end{eqnarray}%
along with%
\begin{eqnarray}
\left[ \tilde{J},\tilde{Q}_{\alpha }^{\pm }\right] &=&-\frac{1}{2}\left(
\gamma _{0}\right) _{\alpha }^{\text{ }\beta }\tilde{Q}_{\beta }^{\pm
}\,,\qquad \ \ \left[ \tilde{J},\tilde{\Sigma}_{\alpha }^{\pm }\right] =-%
\frac{1}{2}\left( \gamma _{0}\right) _{\alpha }^{\text{ }\beta }\tilde{\Sigma%
}_{\beta }^{\pm }\,,  \notag \\
\qquad \left[ \tilde{H},\tilde{Q}_{\alpha }^{-}\right] &=&-\left( \gamma
_{0}\right) _{\alpha }^{\text{ }\beta }\tilde{\Sigma}_{\beta }^{-}\,,\qquad
\quad \left[ \tilde{P}_{a},\tilde{Q}_{\alpha }^{+}\right] =-\frac{1}{2}%
\left( \gamma _{a}\right) _{\alpha }^{\text{ }\beta }\tilde{\Sigma}_{\beta
}^{-}\,,  \notag \\
\left[ \tilde{G}_{a},\tilde{Q}_{\alpha }^{+}\right] &=&-\frac{1}{2}\left(
\gamma _{a}\right) _{\alpha }^{\text{ }\beta }\tilde{Q}_{\beta
}^{-}\,,\qquad \left[ \tilde{G}_{a},\tilde{\Sigma}_{\alpha }^{+}\right] =-%
\frac{1}{2}\left( \gamma _{a}\right) _{\alpha }^{\text{ \ }\beta }\tilde{%
\Sigma}_{\beta }^{-}\,,  \notag \\
\left\{ \tilde{Q}_{\alpha }^{-},\tilde{Q}_{\beta }^{-}\right\} &=&-2\left(
\gamma ^{0}C\right) _{\alpha \beta }\tilde{M}\,,\quad \left\{ \tilde{Q}%
_{\alpha }^{+},\tilde{Q}_{\beta }^{-}\right\} =-\left( \gamma ^{a}C\right)
_{\alpha \beta }\tilde{P}_{a}\,,  \label{N2} \\
\left\{ \tilde{Q}_{\alpha }^{+},\tilde{Q}_{\beta }^{+}\right\} &=&-\left(
\gamma ^{0}C\right) _{\alpha \beta }\tilde{H}\,,\quad \quad \left\{ \tilde{Q}%
_{\alpha }^{-},\tilde{\Sigma}_{\beta }^{-}\right\} =-2\left( \gamma
^{0}C\right) _{\alpha \beta }\tilde{T}\,,  \notag \\
\left\{ \tilde{Q}_{\alpha }^{\pm },\tilde{\Sigma}_{\beta }^{\mp }\right\}
&=&-\left( \gamma ^{a}C\right) _{\alpha \beta }\tilde{Z}_{a}\,,\quad \ \
\left\{ \tilde{Q}_{\alpha }^{+},\tilde{\Sigma}_{\beta }^{+}\right\} =-\left(
\gamma ^{0}C\right) _{\alpha \beta }\tilde{Z}\,.  \notag
\end{eqnarray}

Notice that, unlike the $\mathcal{N}=1$ superalgebra, the $\mathcal{N}=2 $
Maxwellian Bargmann superalgebra obtained here can be seen as a true
supersymmetry algebra. In particular, let us note the presence of the
non-vanishing commutator between the $\tilde{G}_{a}$ generator and
supersymmetry generator. Nevertheless, this superalgebra does not contain
the MEB\ algebra as a subalgebra. Indeed, the bosonic subalgebra (\ref{MB})
can be seen as a non-relativistic version of a [Maxwell]$\oplus u\left(
1\right) \oplus u\left( 1\right) $ algebra.

Moreover, although the $\mathcal{N}=2$ NR Maxwell superalgebra (\ref{MB})-(%
\ref{N2}) has the desired features of a true superalgebra, it is not a good
candidate to construct a three-dimensional CS supergravity action. Indeed,
in order to have a NR supergravity action based on a supersymmetric
extension of the MEB algebra, we need a well-defined invariant tensor, which
requires to introduce by hand additional fermionic generators. The explicit
Maxwellian extended Bargmann superalgebra allowing to construct a NR
supergravity action is presented in the next section.

\section{Maxwellian extended Bargmann supergravity}

Here{,} we present the explicit form of the Maxwellian extended Bargmann
superalgebra allowing to construct a NR supergravity action. Consequently,
we develop the aforementioned NR supergravity action by exploiting the CS
construction in three dimensions.

\subsection{Maxwellian extended Bargmann superalgebra}

As we have discussed in the previous section, the $\mathcal{N}=2$ NR Maxwell
superalgebra given by (\ref{MB})-(\ref{N2}) does not allow for the proper
construction of a NR CS supergravity action although its relativistic
analogue is well-defined.
In order to have a proper NR CS supergravity
action based on a supersymmetric extension of the MEB algebra, one requires
to find a NR superalgebra which not only contains the MEB algebra as a
subalgebra but also admits a non-degenerate invariant supertrace. {Indeed, when studying the NR limit of a theory, one has that the symplectic form of the NR model might become degenerate, making some fields not determined by the field equations, thus reducing the number of dynamical fields. In the case of a CS formulation in three dimensions, the non-degeneracy of the bilinear invariant trace of gauge generators implies the non-degeneracy of the symplectic form, which would ensure dynamically indeterminate fields in the NR theory. In particular, the non-degeneracy of the bilinear form is related to the physical requirement that the CS action involves a kinematical term for each field and the equation of motions imply that all curvatures vanish.}

Here we construct by hand a supersymmetric extension of the MEB algebra by
introducing six Majorana fermionic generators $\tilde{Q}_{{\alpha}}^{+}$, $\tilde{Q}%
_{{\alpha}}^{-} $, $\tilde{\Sigma}_{{\alpha}}^{+}$, $\tilde{\Sigma}_{{\alpha}}^{-}$, $\tilde{R}_{{\alpha}}$, and $\tilde{%
W}_{{\alpha}}$. Let us note that the presence of the $\tilde{R}_{{\alpha}}$ and $\tilde{W}_{{\alpha}}$
generators is similar to what happens in the extended Bargmann superalgebra
presented in \cite{BR} and in the extended Newtonian superalgebra of \cite%
{OOTZ}, in which a $\tilde{R}_{{\alpha}}$ generator is considered. Furthermore, we
introduce six extra bosonic generators $Y_1$, $Y_2$, $U_1$, $U_2$, $B_1$,
and $B_2$. Both $B_1$ and $B_2$ are central, while the others act
non-trivially on the spinor generators, similarly to the extra bosonic
generators introduced in the extended Newton-Hooke supergravity \cite{OOZ}.
The proposed supersymmetric extension of the MEB algebra is generated by the
set of bosonic and fermionic generators
\begin{equation}
\{\tilde{J},\tilde{G}_{a},\tilde{S},\tilde{H},\tilde{P}_{a},\tilde{M},\tilde{%
Z},\tilde{Z}_{a},\tilde{T},\tilde{Y}_{1},\tilde{Y}_{2},\tilde{U}_{1},\tilde{U%
}_{2},\tilde{B}_{1},\tilde{B}_{2},\tilde{Q}_{\alpha }^{+},\tilde{Q}_{\alpha
}^{-},\tilde{R}_{\alpha },\tilde{\Sigma}_{\alpha }^{+},\tilde{\Sigma}%
_{\alpha }^{-},\tilde{W}_{\alpha }\}.
\end{equation}

Such generators satisfy the MEB algebra (\ref{MEB1}) along with the
following non-vanishing (anti-)commutation relations:

{\begin{equation}\label{sMEB}
\begin{split}
& \left[ \tilde{J},\tilde{Q}_{\alpha }^{\pm }\right] =-\frac{1}{2}\left(
\gamma _{0}\right) _{\alpha }^{\text{ }\beta }\tilde{Q}_{\beta }^{\pm
}\,, \quad \left[ \tilde{J},\tilde{\Sigma}_{\alpha }^{\pm }\right] =-%
\frac{1}{2}\left( \gamma _{0}\right) _{\alpha }^{\text{ }\beta }\tilde{\Sigma%
}_{\beta }^{\pm }\,,  \quad \left[ \tilde{M},\tilde{Q}_{\alpha }^{+}\right] = -\frac{1}{2}\left( \gamma
_{0}\right) _{\alpha }^{\text{ }\beta }\tilde{W}_{\beta } \,, \\
& \left[ \tilde{H},\tilde{Q}_{\alpha }^{\pm }\right] =-\frac{1}{2}\left(
\gamma _{0}\right) _{\alpha }^{\text{ }\beta }\tilde{\Sigma}_{\alpha }^{\pm
}\,, \quad \left[ \tilde{P}_{a},\tilde{Q}_{\alpha }^{+}\right] =-\frac{1}{2}\left( \gamma _{a}\right) _{\alpha }^{\text{ }\beta }\tilde{\Sigma}_{\beta}^{-}\,,  \quad \left[ \tilde{Y}_{1},\tilde{Q}_{\alpha }^{\pm}\right] = \pm \frac{1}{2}\left(\gamma _{0}\right) _{\alpha \beta }\tilde{Q}_{\beta }^{\pm} \,, \\
& \left[ \tilde{G}_{a},\tilde{Q}_{\alpha }^{+}\right] = -\frac{1}{2}\left(
\gamma _{a}\right) _{\alpha }^{\text{ }\beta }\tilde{Q}_{\beta}^{-}\,, \quad \left[ \tilde{G}_{a},\tilde{Q}_{\alpha }^{-}\right] =-\frac{1}{2}\left( \gamma _{a}\right) _{\alpha }^{\text{ }\beta }\tilde{R}_{\beta
}\,,  \quad  \left[ \tilde{Y}_{1},\tilde{\Sigma}_{\alpha }^{\pm}\right] = \pm \frac{1}{2}\left(\gamma _{0}\right) _{\alpha \beta }\tilde{\Sigma}_{\beta }^{\pm} \,, \\
& \left[ \tilde{G}_{a},\tilde{\Sigma}_{\alpha }^{+}\right] = -\frac{1}{2}
\left( \gamma _{a}\right) _{\alpha }^{\text{ }\beta }\tilde{\Sigma}_{\beta}^{-}\,, \quad \left[ \tilde{G}_{a},\tilde{\Sigma}_{\alpha }^{-}\right] =-\frac{1}{2}\left( \gamma _{a}\right) _{\alpha }^{\text{ }\beta }%
\tilde{W}_{\beta }\,,  \quad  \left[ \tilde{U}_{1},\tilde{Q}_{\alpha }^{\pm}\right] = \pm \frac{1}{2}\left(\gamma _{0}\right) _{\alpha \beta }\tilde{\Sigma}_{\beta }^{\pm} \,, \\
& \left[ P_{a},\tilde{Q}_{\alpha }^{-}\right] = -\frac{1}{2}\left( \gamma
_{a}\right) _{\alpha }^{\text{ }\beta }\tilde{W}_{\beta }\,,  \ \left[ \tilde{J},\tilde{R}_{\alpha }\right] =-\frac{1}{2}\left( \gamma
_{0}\right) _{\alpha }^{\text{ }\beta }\tilde{R}_{\beta }\,\,, \ \left[ \tilde{Y}_{2},\tilde{Q}_{\alpha }^{+}\right] = \left[ \tilde{Y}_{1},\tilde{R}_{\alpha } \right] =\frac{1}{2}\left(
\gamma _{0}\right) _{\alpha \beta }\tilde{R}_{\beta } \, , \\
& \left[ \tilde{J},\tilde{W}_{\alpha }\right] = -\frac{1}{2}\left( \gamma
_{0}\right) _{\alpha }^{\text{ }\beta }\tilde{W}_{\beta }\,,  \
\left[ \tilde{H},\tilde{R}_{\alpha }\right] =-\frac{1}{2}\left( \gamma
_{0}\right) _{\alpha }^{\text{ }\beta }\tilde{W}_{\beta }\,,  \   \left[ \tilde{Y}_{2},\tilde{\Sigma}_{\alpha }^{+}\right] = \left[ \tilde{Y}_{1},\tilde{W}_{\alpha } \right]
=\frac{1}{2}\left( \gamma _{0}\right) _{\alpha \beta }\tilde{W}_{\beta } \,, \\
& \left[ \tilde{S},\tilde{Q}_{\alpha }^{+}\right] = -\frac{1}{2}\left( \gamma
_{0}\right) _{\alpha }^{\text{ }\beta }\tilde{R}_{\beta }\,,  \
\left[ \tilde{S},\tilde{\Sigma}_{\alpha }^{+}\right] =-\frac{1}{2}\left(
\gamma _{0}\right) _{\alpha }^{\text{ }\beta }\tilde{W}_{\beta }\,,   \ \left[ \tilde{U}_{2},\tilde{Q}_{\alpha }^{+}\right] = \left[ \tilde{U}_{1},\tilde{R}_{\alpha } \right] = \frac{1}{2}\left( \gamma _{0}\right) _{\alpha \beta }\tilde{W}_{\beta } \,,  \\
& \left\{ \tilde{Q}_{\alpha }^{-},\tilde{Q}_{\beta }^{-}\right\} = -\left(
\gamma ^{0}C\right) _{\alpha \beta }\tilde{M}{+}\left( \gamma ^{0}C\right)
_{\alpha \beta }\tilde{U}_{2}\,, \quad  \left\{ \tilde{Q}_{\alpha }^{+},\tilde{\Sigma}_{\beta }^{+}\right\}
=-\left( \gamma ^{0}C\right) _{\alpha \beta }\tilde{Z}-\left( \gamma
^{0}C\right) _{\alpha \beta }\tilde{B}_{1} \,,  \\
& \left\{ \tilde{Q}_{\alpha }^{+},\tilde{Q}_{\beta }^{+}\right\} = -\left(
\gamma ^{0}C\right) _{\alpha \beta }\tilde{H}-\left( \gamma ^{0}C\right)
_{\alpha \beta }\tilde{U}_{1}\,,  \quad \left\{ \tilde{Q}_{\alpha }^{-},\tilde{\Sigma}_{\beta }^{-}\right\}
= -\left( \gamma ^{0}C\right) _{\alpha \beta }\tilde{T}{+}\left( \gamma
^{0}C\right) _{\alpha \beta }\tilde{B}_{2}\,,  \\
& \left\{ \tilde{Q}_{\alpha }^{+},\tilde{R}_{\beta }\right\} = -\left( \gamma
^{0}C\right) _{\alpha \beta }\tilde{M}-\left( \gamma ^{0}C\right) _{\alpha
\beta }\tilde{U}_{2}\,, \quad \left\{ \tilde{Q}_{\alpha }^{+},\tilde{W}_{\beta }\right\} =-\left( \gamma
^{0}C\right) _{\alpha \beta }\tilde{T}-\left( \gamma ^{0}C\right) _{\alpha
\beta }\tilde{B}_{2}\,,  \\
& \left\{ \tilde{\Sigma}_{\alpha }^{+},\tilde{R}_{\beta }\right\} = -\left(
\gamma ^{0}C\right) _{\alpha \beta }\tilde{T}-\left( \gamma ^{0}C\right)
_{\alpha \beta }\tilde{B}_{2} \,, \quad \ \left\{ \tilde{Q}_{\alpha }^{\pm },\tilde{\Sigma}_{\beta }^{\mp }\right\} = -\left( \gamma ^{a}C\right) _{\alpha \beta }\tilde{Z}_{a}\,, \\
& \left\{ \tilde{Q}_{\alpha }^{+},\tilde{Q}_{\beta }^{-}\right\} =-\left(\gamma ^{a}C\right) _{\alpha \beta }\tilde{P}_{a} \,.
\end{split}
\end{equation}}

The superalgebra given by {(\ref{MEB1}) and (\ref{sMEB})} will be denoted as the Maxwellian extended Bargmann
superalgebra. One can note that the $\tilde{S}$ generator is no longer a
central charge in this supersymmetric extension of the MEB algebra {but acts
non-trivially on the spinor generators $\tilde{Q}_{{\alpha}}^{+}$ and $\tilde{\Sigma}_{{\alpha}}^{+}$.} It is important to emphasize that the MEB superalgebra obtained here
has not been obtained through a NR limit of a relativistic superalgebra.
Furthermore, the supersymmetric extension of the MEB algebra allowing a
well-defined CS supergravity action could not be unique. Then, it would be
interesting to study further supersymmetric extensions of the MEB algebra
and the possibility of obtaining them by applying a NR limit to a
relativistic theory.

\subsection{Non-relativistic Chern-Simons supergravity action}

Let us construct a NR CS supergravity action based on the MEB superalgebra {%
previously introduced}.

The non-vanishing components of the invariant tensor for the MEB
superalgebra are given by (\ref{invt1}) along with
\begin{eqnarray}
\left\langle \tilde{Z}\tilde{S}\right\rangle &=&-\tilde{\alpha}_{2}\,,
\notag \\
\left\langle \tilde{Y}_{1}\tilde{Y}_{2}\right\rangle &=&\tilde{\alpha}_{0}\,,
\notag \\
\left\langle \tilde{Y}_{1}\tilde{U}_{2}\right\rangle &=&\tilde{\alpha}%
_{1}\,=\,\left\langle \tilde{U}_{1}\tilde{Y}_{2}\right\rangle \,,  \notag \\
\left\langle \tilde{Y}_{1}\tilde{B}_{2}\right\rangle &=&\tilde{\alpha}%
_{2}\,=\,\left\langle \tilde{U}_{1}\tilde{U}_{2}\right\rangle
\,=\,\left\langle \tilde{B}_{1}\tilde{Y}_{2}\right\rangle \,,  \label{invt2p} \\
\left\langle \tilde{Q}_{\alpha }^{-}\tilde{Q}_{\beta }^{-}\right\rangle &=&2%
\tilde{\alpha}_{1}C_{\alpha \beta }\,=\,\left\langle \tilde{Q}_{\alpha }^{+}%
\tilde{R}_{\beta }\right\rangle \,,  \notag \\
\left\langle \tilde{Q}_{\alpha }^{-}\tilde{\Sigma}_{\beta }^{-}\right\rangle
&=&2\tilde{\alpha}_{2}C_{\alpha \beta }\,=\,\left\langle \tilde{\Sigma}%
_{\alpha }^{+}\tilde{R}_{\beta }\right\rangle \,=\,\left\langle \tilde{Q}%
_{\alpha }^{+}\tilde{W}_{\beta }\right\rangle \,,  \notag
\end{eqnarray}%
where $\tilde{\alpha}_{0}$, $\tilde{\alpha}_{1}$, and $\tilde{\alpha}_{2}$
are arbitrary constants. The bilinear form associated with the MEB
superalgebra is non-degenerate for $\tilde{\alpha} _{2}\neq 0$, analogously
to the purely bosonic case \cite{AFGHZ}. {On the other hand,} the gauge
connection one-form $\tilde{A}$ for the MEB superalgebra reads\footnote{{Here and in the sequel, we omit the spinor index $\alpha$, in order to lighten the notation.}}
\begin{eqnarray}
\tilde{A} &=&\omega \tilde{J}+\omega ^{a}\tilde{G}_{a}+\tau \tilde{H}+e%
\tilde{P}_{a}+k\tilde{Z}+k^{a}\tilde{Z}_{a}+m\tilde{M}+s\tilde{S}+t\tilde{T}
+y_{1}\tilde{Y_{1}}+y_{2}\tilde{Y}_{2}+b1\tilde{B}_{1}\notag \\
&&+b_{2}\tilde{B}%
_{2}+u_{1}\tilde{U}_{1}+u_{2}\tilde{U}_{2}+{\psi }^{+}\tilde{Q}^{+}+{\psi }^{-}\tilde{Q}^{-}+{\xi }^{+}\tilde{\Sigma}%
^{+}+{\xi }^{-}\tilde{\Sigma}^{-}+{\rho }\tilde{R}+{\chi }\tilde{W}\,\,.
\label{oneform2}
\end{eqnarray}%
The corresponding curvature two-form $\tilde{F}=d\tilde{A}+\tilde{A}\wedge
\tilde{A}=d\tilde{A}+\frac{1}{2}\left[ \tilde{A},\tilde{A}\right] $ in terms
of the generators is given by
\begin{eqnarray}
\tilde{F} &=&R\left( \omega \right) \tilde{J}+R^{a}\left( \omega ^{b}\right)
\tilde{G}_{a}+F\left( \tau \right) \tilde{H}+F^{a}\left( e^{b}\right) \tilde{%
P}_{a}+F\left( k\right) \tilde{Z}+F^{a}\left( k^{b}\right) \tilde{Z}%
_{a}+F\left( m\right) \tilde{M}  \notag \\
&&+R\left( s\right) \tilde{S}+F\left( t\right) \tilde{T}+F\left(
y_{1}\right) \tilde{Y_{1}}+F\left( y_{2}\right) \tilde{Y}_{2}+F\left(
b_{1}\right) \tilde{B}_{1}+F\left( b_{2}\right) \tilde{B}_{2}+F\left(
u_{1}\right) \tilde{U}_{1}  \notag \\
&&+F\left( u_{2}\right) \tilde{U}_{2}+\nabla {\psi }^{+}\tilde{Q}^{+}+\nabla
{\psi }^{-}\tilde{Q}^{-}+\nabla {\xi }^{+}\tilde{\Sigma}^{+}+\nabla {\xi }%
^{-}\tilde{\Sigma}^{-}+\nabla {\rho }\tilde{R}+\nabla {\chi }\tilde{W}\,.
\label{F2c}
\end{eqnarray}%
{In particular,} the bosonic curvature two-forms are given by
\begin{eqnarray}
F\left( y_{1}\right) &=&dy_{1}\,,  \notag \\
F\left( y_{2}\right) &=&dy_{2}\,,  \notag \\
F\left( b_{1}\right) &=&db_{1}+\bar{\psi}^{+}\gamma ^{0}\xi ^{+}\,,  \notag
\\
F\left( b_{2}\right) &=&db_{2}-\bar{\psi}^{-}\gamma ^{0}\xi ^{-}+\bar{\psi}%
^{+}\gamma ^{0}\chi +\bar{\xi}^{+}\gamma ^{0}\rho \,,
\label{boscurvSuperMEBp2} \\
F\left( u_{1}\right) &=&du_{1}+\frac{1}{2}\bar{\psi}^{+}\gamma ^{0}\psi
^{+}\,,  \notag \\
F\left( u_{2}\right) &=&du_{2}-\frac{1}{2}\bar{\psi}^{-}\gamma ^{0}\psi ^{-}+%
\bar{\psi}^{+}\gamma ^{0}\rho \,.  \notag
\end{eqnarray}%
together with
\begin{eqnarray}
F\left( \omega \right) &=&R\left( \omega \right) \,,  \notag \\
F^{a}\left( \omega ^{b}\right) &=&R^{a}\left( \omega ^{b}\right) \,,  \notag
\\
F\left( \tau \right) &=&R\left( \tau \right) +\frac{1}{2}\bar{\psi}%
^{+}\gamma ^{0}\psi ^{+}\,,  \notag \\
F^{a}\left( e^{b}\right) &=&R^{a}\left( e^{b}\right) +\bar{\psi}^{+}\gamma
^{a}\psi ^{-}\,,  \notag \\
F\left( k\right) &=&R\left( k\right) +\bar{\psi}^{+}\gamma ^{0}\xi ^{+}\,,
\label{boscurvSuperMEBp1} \\
F^{a}\left( k^{b}\right) &=&R^{a}\left( k^{b}\right) +\bar{\psi}^{+}\gamma
^{a}\xi ^{-}+\bar{\psi}^{-}\gamma ^{a}\xi ^{+}\,,  \notag \\
F\left( m\right) &=&R\left( m\right) +\frac{1}{2}\bar{\psi}^{-}\gamma
^{0}\psi ^{-}+\bar{\psi}^{+}\gamma ^{0}\rho \,,  \notag \\
F\left( s\right) &=&R\left( s\right) \,,  \notag \\
F\left( t\right) &=&R\left( t\right) +\bar{\psi}^{-}\gamma ^{0}\xi ^{-}+\bar{%
\psi}^{+}\gamma ^{0}\chi +\bar{\xi}^{+}\gamma ^{0}\rho \,,  \notag
\end{eqnarray}%
where $R\left( \omega \right) $, $R^{a}\left( \omega ^{b}\right) $, $R\left(
\tau \right) $, $R^{a}\left( e^{b}\right) $, $R\left( k\right) $, $%
R^{a}\left( k^{b}\right) $, $R\left( m\right) $, $R\left( s\right) $, and $%
R\left( t\right) $ correspond to the MEB curvatures already defined in (\ref{curvMEB}). On the other hand, the covariant derivatives of the spinor $1$-form fields
read
\begin{eqnarray}
\nabla \psi ^{+} &=&d\psi ^{+}+\frac{1}{2}\omega \gamma _{0}\psi ^{+}-\frac{1%
}{2}y_{1}\gamma _{0}\psi ^{+}\,,  \notag \\
\nabla \psi ^{-} &=&d\psi ^{-}+\frac{1}{2}\omega \gamma _{0}\psi ^{-}+\frac{1%
}{2}\omega ^{a}\gamma _{a}\psi ^{+}+\frac{1}{2}y_{1}\gamma _{0}\psi ^{-}\,,
\notag \\
\nabla \xi ^{+} &=&d\xi ^{+}+\frac{1}{2}\omega \gamma _{0}\xi ^{+}+\frac{1}{2%
}\tau \gamma _{0}\psi ^{+}-\frac{1}{2}y_{1}\gamma _{0}\xi ^{+}-\frac{1}{2}%
u_{1}\gamma _{0}\psi ^{+}\,,  \notag \\
\nabla \xi ^{-} &=&d\xi ^{-}+\frac{1}{2}\omega \gamma _{0}\xi ^{-}+\frac{1}{2%
}\tau \gamma _{0}\psi ^{-}+\frac{1}{2}e^{a}\gamma _{a}\psi ^{+}+\frac{1}{2}%
\omega ^{a}\gamma _{a}\xi ^{+}  \label{fermcurvSuperMEB} \\
&&+\frac{1}{2}y_{1}\gamma _{0}\xi ^{-}+\frac{1}{2}u_{1}\gamma _{0}\psi
^{-}\,,  \notag \\
\nabla \rho &=&d\rho +\frac{1}{2}\omega \gamma _{0}\rho +\frac{1}{2}\omega
^{a}\gamma _{a}\psi ^{-}+\frac{1}{2}s\gamma _{0}\psi ^{+}-\frac{1}{2}%
y_{2}\gamma _{0}\psi ^{+}-\frac{1}{2}y_{1}\gamma _{0}\rho \,,  \notag \\
\nabla \chi &=&d\chi +\frac{1}{2}\omega \gamma _{0}\chi +\frac{1}{2}\omega
^{a}\gamma _{a}\xi ^{-}+\frac{1}{2}e^{a}\gamma _{a}\psi ^{-}+\frac{1}{2}\tau
\gamma _{0}\rho +\frac{1}{2}s\gamma _{0}\xi ^{+}+\frac{1}{2}m\gamma _{0}\psi
^{+}  \notag \\
&&-\frac{1}{2}y_{2}\gamma _{0}\xi ^{+}-\frac{1}{2}y_{1}\gamma _{0}\chi -%
\frac{1}{2}u_{2}\gamma _{0}\psi ^{+}-\frac{1}{2}u_{1}\gamma _{0}\rho \,.
\notag
\end{eqnarray}%
Let us note that the gauge fields related to the additional bosonic generators $\{Y_{1},Y_{2},U_{1},U_{2}\}$ appear explicitly on the spinorial curvature. One can see that $b_1$ and $b_2$ do not give any contribution to the fermionic curvatures since they are related to central charges.

A CS supergravity action based on the Maxwellian extended Bargmann superalgebra can be constructed
by combining the non-zero invariant tensors (\ref{invt1}) and (\ref{invt2p}) with the gauge connection one-form $\tilde{A}$ (\ref%
{oneform2}), and it reads, up to boundary terms, as follows:
\begin{eqnarray}
I &=&\int \Bigg \lbrace \tilde{\alpha}_{0} \bigg[ \omega _{a}R^{a}(\omega
^{b})-2sR\left(\omega \right) +2y_{1}dy_{2}\bigg] +\tilde{\alpha}_{1} \bigg[ %
2e_{a}R^{a}(\omega ^{b})-2mR(\omega )-2\tau R(s)+2y_{1}du_{2}  \notag \\
&& +2u_{1}dy_{2}+2\bar{\psi}^{+}\nabla \rho +2\bar{\rho}\nabla \psi ^{+}+2%
\bar{\psi}^{-}\nabla \psi ^{-} \bigg] +\tilde{\alpha}_{2} \bigg[ %
e_{a}R^{a}\left( e^{b}\right) +k_{a}R^{a}\left( \omega ^{b}\right) +\omega
_{a}R^{a}\left( k^{b}\right)  \notag \\
&& -2sR\left( k\right) -2mR\left( \tau \right) -2tR\left( \omega \right)
+2y_{1}db_{2}+2u_{1}du_{2}+2y_{2}db_{1}+2\bar{\psi}^{-}\nabla \xi ^{-}+2\bar{%
\xi}^{-}\nabla \psi ^{-}  \notag \\
&& +2\bar{\psi}^{+}\nabla \chi +2\bar{\chi}\nabla \psi ^{+}+2\bar{\xi}%
^{+}\nabla \rho +2\bar{\rho}\nabla \xi ^{+} \bigg] \Bigg \rbrace \,.
\label{CS2}
\end{eqnarray}%
The CS action (\ref{CS2}) obtained here describes the so-called Maxwellian
extended Bargmann supergravity theory. Let us note that the NR CS
supergravity action (\ref{CS2}) contains three independent sectors
proportional to $\tilde{\alpha}_{0}$, $\tilde{\alpha}_{1}$, and $\tilde{%
\alpha}_{2}$. In particular, the term proportional to $\tilde{\alpha}_{0}$
corresponds to a NR exotic Lagrangian, while the extended Bargmann
supergravity introduced in \cite{BR} appears in the $\tilde{\alpha}_{1}$
sector, endowed with some additional terms related to the presence of the
bosonic $1$-form fields $y_{1}$, $y_{2}$, $u_{1}$, and $u_{2}$. The term
proportional to $\tilde{\alpha}_{2}$ can be seen as the CS Lagrangian for a
new NR Maxwell superalgebra.
In addition, one can see that the bosonic part of the CS action (\ref{CS2})
corresponds to the MEB gravity action presented in \cite{AFGHZ},
supplemented with the bosonic $1$-form fields $y_{1}$, $y_{2}$, $b_{1}$, $%
b_{2}$, $u_{1}$, and $u_{2}$.

Note that, for $\alpha _{2}\neq 0$, the field equations from the NR CS
supergravity action (\ref{CS2}) reduce to the vanishing of the curvature
two-forms (\ref{boscurvSuperMEBp1}), (\ref{boscurvSuperMEBp2}), and (\ref%
{fermcurvSuperMEB}) associated with the {MEB} superalgebra.
{These curvatures transform covariantly with respect to the supersymmetry transformation laws given in (\ref{GT}).}

The three-dimensional Maxwellian extended Bargmann supergravity theory
obtained here corresponds to an alternative NR supergravity theory which
contains the extended Bargmann supergravity \cite{BR} (supplemented with
some additional bosonic $1$-form fields) as a sub-case and which is distinct
from the {Newton-Cartan} supergravity introduced in \cite{ABRS, BRZ}. {It is important to emphasize that, although our result generalizes the extended Bargmann supergravity \cite{BR}, the new NR supergravity obtained here do not contain a cosmological constant. It would be then interesting to study the introduction of a cosmological constant in our model.}

\section{Discussion}

In this work{,} we have studied the NR limit of the relativistic Maxwell
superalgebra. A well-defined NR superalgebra with the desired features has
been obtained by contracting the $\mathcal{N}=2$ Maxwell superalgebra
introduced in \cite{Concha}. Nevertheless, the construction of a proper NR
supergravity action based on a NR version of the Maxwell superalgebra has
required to introduce by hand new fermionic and bosonic generators. The new
structure has been called as the Maxwellian extended Bargmann superalgebra
and corresponds to a supersymmetric extension of the MEB algebra presented
in \cite{AFGHZ}. In particular the MEB\ superalgebra admits a non-degenerate
invariant bilinear form allowing to construct a proper NR CS supergravity
action. Interestingly, the MEB CS supergravity theory presented here
contains the extended Bargmann supergravity as a sub-case.

{The NR supergravity action constructed in this work, see (\ref{CS2}), could serve as a starting point for diverse studies. In particular, our result is the first step to construct a new family of NR supergravity beyond the extended Bargmann supergravity. Furthermore,} the MEB supergravity could
be useful in the construction of a three-dimensional Horava-Lifshitz
supergravity. Indeed, as was noticed in \cite{BR, HLO}, the extended
Bargmann gravity can be seen as a particular kinetic term of the
Horava-Lifshitz gravity. In particular, it would be intriguing to explore
the effects arising from the presence of the additional gauge field
appearing in the Maxwell version of the extended Bargmann (super)algebra.

{Moreover, the NR supergravity action (\ref{CS2}) could have applications in the context of holography, as it already happens for NR gravity at the purely bosonic level \cite{BG, CHOR}, as well as in effective field theory, where gravitational fields are used as geometrical background response functions \cite{Son2, Geracie:2014nka}. Also at this level, it would be interesting to examine the role played by the additional field involved in Maxwell version of the extended Bargmann (super)algebra. As a further remark, let us highlight that additional gauge fields appearing in (super)Maxwell algebras have already been proven to be of particular relevance in the study of the supersymmetry invariance of flat (relativistic) supergravity on a manifold with non-trivial boundary \cite{CRR}, where it has also been conjectured that the presence of the new gauge fields in the boundary would allow to regularize the supergravity action in the holographic renormalization language.}

{On the other hand, concerning, in particular, the purely bosonic restriction of our model, one could explore the phenomenological implications possibly related to Dark Matter at the non-relativistic level. Another aspect of the MEB gravity which deserves some investigation is the study of the asymptotic symmetries. As it was shown in \cite{CMMRSV}, the asymptotic symmetry of the three-dimensional CS theory invariant under the Maxwell algebra is given by an extension and deformation of the BMS$_3$ algebra denoted as $\mathfrak{Max}_3$ \cite{CCSS}. One could then expect that, since the MEB gravity is the NR version of the Maxwell CS gravity, the corresponding asymptotic symmetry of the MEB gravity would be given by a NR version of the $\mathfrak{Max}_3$ algebra. Interestingly, the presence of the additional gauge field $Z_A$ modifies not only the asymptotic sector but also the vacuum of the theory. It would be then interesting to study the physical implications of the additional bosonic gauge field appearing in our model. One could go even further and extend this study to supersymmetric level. However, this would require first to establish the relativistic version of our NR model and then study its asymptotic structure by considering suitable boundary conditions.}

{It would also be interesting to develop higher-dimensional extensions of our NR supergravity theory, since we argue that, due to the fact that the introduction of (at least) one second
spinorial charge not only is required to construct the minimal supersymmetric extension of the Maxwell algebra but has also been considered in superstring \cite{Green} and in $D=11$ supergravity \cite{AF, AAR1, AAR2}, they could be relevant in the study and analysis of the NR limit of higher-dimensional supergravity models, and, more specifically, of $D=11$ supergravity.}

A future development could also consist in exploring the possibility to
obtain the MEB superalgebra introduced here through a NR limit or
contraction process from a relativistic superalgebra. An alternative limit
which could be used to recover the MEB superalgebra is the vanishing cosmological
constant limit. In particular, {it would be worth exploring the possibility to accommodate a cosmological constant to the MEB supergravity presented here.} One could conjecture that a supersymmetric
extension of the recent enlarged extended Bargmann gravity, introduced in
\cite{CR4}, reproduces the present MEB supergravity in a flat limit [work in
progress].

On the other hand, it would be interesting to apply the Lie algebra
expansion method \cite{HSa, AIPV, AIPV2, Sexp} to obtain the MEB
superalgebra. One could follow the procedure used in \cite{BIOR, AGI, Romano}
and consider the expansion of a relativistic Maxwell superalgebra.
Alternatively, one might also extend the results obtained in \cite{CR4, MPS}
in which NR algebras appear as semigroup expansions of the so-called
Nappi-Witten algebra.

Another aspect that deserves further investigation regards the development
of a Maxwellian version of the extended Newtonian gravity \cite{HHO} and its
supersymmetric extension \cite{OOTZ}. {In particular, the result obtained here could correspond to a subcase of a Maxwellian generalization of the extended Newtonian supergravity.} A cosmological constant has recently
been accommodated in the extended Newtonian gravity action by including new
generators to the Newton-Hooke algebra \cite{CRR2}. One could expect to
obtain a Maxwellian Newtonian algebra by generalizing the MEB one in a
similar way to \cite{OOTZ, CRR2}. It would be then compelling to explore
possible matter couplings.

\section*{Acknowledgment}

This work was supported by the CONICYT - PAI grant N$^{\circ }$77190078
(P.C.) and FONDECYT Projects N$^{\circ }$3170438 (E.R.). P.C. would like to thank to the Dirección de Investigación and Vice-rectoría de Investigación of the Universidad Católica de la Santísima Concepción, Chile, for their constant support.

\appendix

\section{{Supersymmetry transformation laws}}

{The curvatures (\ref{boscurvSuperMEBp1}), (\ref{boscurvSuperMEBp2}), and (\ref{fermcurvSuperMEB}) associated with the MEB superalgebra} transform covariantly with respect to the following supersymmetry transformation laws:
\begin{eqnarray}
\delta \omega &=&0\,,  \notag \\
\delta \omega ^{a} &=&0\,,  \notag \\
\delta \tau &=&-\bar{\epsilon}^{+}\gamma ^{0}\psi ^{+}\,,  \notag \\
\delta e^{a} &=&-\bar{\epsilon}^{+}\gamma ^{a}\psi ^{-}-\bar{\epsilon}%
^{-}\gamma ^{a}\psi ^{+}\,,  \notag \\
\delta k &=&-\bar{\epsilon}^{+}\gamma ^{0}\xi ^{+}-\bar{\varphi}^{+}\gamma
^{0}\psi ^{+}\,, \notag \\
\delta k^{a} &=&-\bar{\epsilon}^{\pm }\gamma ^{a}\xi ^{\mp }-\bar{\varphi}%
^{\pm }\gamma ^{a}\psi ^{\mp }\,,  \notag \\
\delta m &=&-\bar{\epsilon}^{-}\gamma ^{0}\psi ^{-}-\bar{\epsilon}^{+}\gamma
^{0}\rho -\bar{\eta}\gamma ^{0}\psi ^{+}\,,  \notag \\
\delta s &=&0\,,  \notag \\
\delta t &=&-\bar{\epsilon}^{-}\gamma ^{0}\xi ^{-}-\bar{\varphi}^{-}\gamma
^{0}\psi ^{-}-\bar{\epsilon}^{+}\gamma ^{0}\chi -\bar{\zeta}\gamma ^{0}\psi
^{+}-\bar{\varphi}^{+}\gamma ^{0}\rho -\bar{\eta}\gamma ^{0}\xi ^{+}\,,
\notag \\
\delta y_{1} &=&0\,,  \notag \\
\delta y_{2} &=&0\,,  \notag \\
\delta b_{1} &=&-\bar{\epsilon}^{+}\gamma ^{0}\xi ^{+}-\bar{\varphi}%
^{+}\gamma ^{0}\psi ^{+}\,,  \notag \\
\delta b_{2} &=&\bar{\epsilon}^{-}\gamma ^{0}\xi ^{-}+\bar{\varphi}%
^{-}\gamma ^{0}\psi ^{-}-\bar{\epsilon}^{+}\gamma ^{0}\chi -\bar{\zeta}%
\gamma ^{0}\psi ^{+}-\bar{\varphi}^{+}\gamma ^{0}\rho -\bar{\eta}\gamma
^{0}\xi ^{+}\,,  \notag \\
\delta u_{1} &=&-\bar{\epsilon}^{+}\gamma ^{0}\psi ^{+}\,,  \notag \\
\delta u_{2} &=&\bar{\epsilon}^{-}\gamma ^{0}\psi ^{-}-\bar{\epsilon}%
^{+}\gamma ^{0}\rho -\bar{\eta}\gamma ^{0}\psi ^{+}\,,  \notag \\
\delta \epsilon ^{+} &=&d\epsilon ^{+}+\frac{1}{2}\omega \gamma _{0}\epsilon
^{+}-\frac{1}{2}y_{1}\gamma _{0}\epsilon ^{+}\,,  \notag \\
\delta \epsilon ^{-} &=&d\epsilon ^{-}+\frac{1}{2}\omega \gamma _{0}\epsilon
^{-}+\frac{1}{2}\omega ^{a}\gamma _{a}\epsilon ^{+}+\frac{1}{2}y_{1}\gamma
_{0}\epsilon ^{-}\,,  \notag \\
\delta \varphi ^{+} &=&d\varphi ^{+}+\frac{1}{2}\omega \gamma _{0}\varphi
^{+}+\frac{1}{2}\tau \gamma _{0}\epsilon ^{+}-\frac{1}{2}y_{1}\gamma
_{0}\varphi ^{+}-\frac{1}{2}u_{1}\gamma _{0}\epsilon ^{+}\,,  \notag \\
\delta \varphi ^{-} &=&d\varphi ^{-}+\frac{1}{2}\omega \gamma _{0}\varphi
^{-}+\frac{1}{2}\tau \gamma _{0}\epsilon ^{-}+\frac{1}{2}e^{a}\gamma
_{a}\epsilon ^{+}+\frac{1}{2}\omega ^{a}\gamma _{a}\varphi ^{+}+\frac{1}{2}%
y_{1}\gamma _{0}\varphi ^{-}+\frac{1}{2}u_{1}\gamma _{0}\epsilon ^{-}\,,
\notag \\
\delta \eta &=&d\eta +\frac{1}{2}\omega \gamma _{0}\eta +\frac{1}{2}\omega
^{a}\gamma _{a}\epsilon ^{-}+\frac{1}{2}s\gamma _{0}\epsilon ^{+}-\frac{1}{2}%
y_{2}\gamma _{0}\epsilon ^{+}-\frac{1}{2}y_{1}\gamma _{0}\eta \,,  \notag \\
\delta \zeta &=&d\zeta +\frac{1}{2}\omega \gamma _{0}\zeta +\frac{1}{2}%
\omega ^{a}\gamma _{a}\varphi ^{-}+\frac{1}{2}e^{a}\gamma _{a}\epsilon ^{-}+%
\frac{1}{2}\tau \gamma _{0}\eta +\frac{1}{2}s\gamma _{0}\varphi ^{+}+\frac{1%
}{2}m\gamma _{0}\epsilon ^{+}  \notag \\
&&-\frac{1}{2}y_{2}\gamma _{0}\varphi ^{+}-\frac{1}{2}y_{1}\gamma _{0}\zeta -%
\frac{1}{2}u_{2}\gamma _{0}\epsilon ^{+}-\frac{1}{2}u_{1}\gamma _{0}\eta \,,
\label{GT}
\end{eqnarray}%
where $\epsilon ^{\pm }$, $\varphi ^{\pm }$, $\eta $, and $\zeta $ are the
fermionic gauge parameter{s} related to {the Majorana fermionic generators} $\tilde{Q}^{\pm }$, $\tilde{\Sigma}^{\pm }$, $\tilde{R}$, and $\tilde{W}$, respectively {(in order to lighten the notation, we have omitted the spinor index $\alpha$)}.

\end{document}